\documentclass[reprint,aps,prl,groupedaddress]{revtex4-1}
\usepackage[margin=0.5in]{geometry} 
\usepackage{graphicx}
\usepackage{dcolumn}
\usepackage{bm}
\usepackage[mathlines]{lineno}
\usepackage{natbib}
\usepackage{hyperref}
\usepackage{relsize}
\usepackage{amsmath,amssymb,amsthm}
\usepackage[usenames,dvipsnames,svgnames,table]{xcolor}
\usepackage{comment}
\usepackage{braket}
\usepackage{url}
\begin{document}
\author{Shah Saad Alam}
\email{shah.saad.alam@gmail.com}
\author{Timothy Skaras}
\author{Li Yang}
\author{Han Pu}
\affiliation{Department of Physics and Astronomy \& Rice Center for Quantum Materials, Rice University, Houston, Texas 77005, USA}
\title{Dynamical Fermionization in One Dimensional Spinor Quantum Gases}

\date{\today}

\begin{abstract}
    Dynamical fermionization refers to the phenomenon in Tonks-Girardeau (TG) gases where, upon release from harmonic confinement, the gas's momentum density profile evolves asymptotically to that of an ideal Fermi gas in the initial trap. This phenomenon has been demonstrated theoretically in hardcore and anyonic TG gases, and recently experimentally observed in a strongly interacting Bose gas. We extend this study to a one dimensional (1D) spinor gas of arbitrary spin in the strongly interacting regime, and analytically prove that the total momentum distribution after the harmonic trap is turned off approaches that of a spinless ideal Fermi gas, while the asymptotic momentum distribution of each spin component takes the same shape of the initial real space density profile of that spin component. Our work demonstrates the rich physics arising from the interplay between the spin and the charge degrees of freedom in a spinor system.
\end{abstract}

\maketitle

Low dimensional systems provide a rich terrain for studying quantum mechanics, since many-body quantum effects are enhanced in reduced dimensions \cite{giamarchi_quantum_2003}. Furthermore, exactly solvable quantum many-body models are rare and precious, and most of such exactly solvable models occur in one dimension (1D) with connections to quantum integrability. The existence of integrability prevents the system from thermalizing, and leads to intriguing quantum dynamics. Together with the experimental accessibility of such systems, this has led to a plenitude of recent attention to such 1D systems.
One example is the Lieb-Liniger model that describes a system of spinless bosons in 1D, and whose properties have been extensively studied in the literature \cite{RigolReviewLieb, Guan2015_CPB}. Adding an inhomogeneous trapping potential generally breaks solvability and quantum integrability. In the infinite interaction (i.e., hardcore) limit, however, it can be solved with the technique of Girardeau's Bose-Fermi mapping \cite{girardeau_relationship_2004}. This extension of Lieb-Liniger gas, named Tonks-Girardeau gas, possesses the same energy spectrum as well as the same real space density profile of an ideal Fermi gas in the same trapping potential. This phenomenon is referred to as {\em fermionization}. It should be pointed out that the Tonks-Girardeau gas and the ideal Fermi gas feature very different momentum density profiles \cite{Girardeau_OPCorrDis_2001, Pezer_BraggTG_2007}.

When a Tonks-Girardeau gas initially confined in a harmonic trap is suddenly released from the trapping potential, the momentum distribution will asymptotically approach that of an ideal spinless Fermi gas in the original trap \cite{minguzzi_exact_2005, Rigol2005DF, Rigol2017Lattice}. This phenomenon has been called {\em dynamical fermionization} (DF). Very recently, dynamical fermionization has been observed in experiment \cite{wilson_observation_2020}. Theoretically, hardcore 1D spinless anyonic gas has also been shown to exhibit DF \cite{del_campo_fermionization_2008}.

Over the past few years, strongly interacting 1D spinor quantum gas has received much attention \cite{Deuretzbacher_QMLattice_2014, volosniev_strongly_2014, Volosniev_2015, levinsen_strong-coupling_2015}. Spinor gases provide a rich playground for quantum magnetism. The additional spin degrees of freedom greatly enrich the physics but, in the same time, also greatly complicates the theoretical investigation. Previous studies have clearly established the fermionization effect in 1D spinor gas in the hardcore limit, but a theory of DF for a general spinor gas is still lacking although studies have been carried out for some special cases~{\cite{Rigol2019}}. In the current work, we fill this gap by investigating the phenomenon of DF for a general hardcore spinor gas initially confined in a harmonic trap. When the initial harmonic trapping potential is suddenly turned off, we analytically prove the following properties in the asymptotic limit:
\begin{itemize}
    \item \textbf{Property 1:} The total momentum distribution (summed over all spin components) approaches that of an ideal spinless Fermi gas in the original trap.
    \item \textbf{Property 2:} The momentum distribution of a spin component takes the same shape as its initial real space distribution.
\end{itemize}
Property 1 represents the analog of DF in the hardcore spinor gas, while Property 2 is a manifestation of the richer physics brought by the spin degrees of freedom. We emphasize that these two properties are valid for both bosons and fermions, and for arbitrary spin. In the following, we will provide a detailed analytic proof of these properties.

{\em Model and wave function in the hardcore limit ---} We consider a 1D harmonically trapped spinor gas with total atom number $N$, either bosonic or fermionic, subject to a two-body $s$-wave contact interaction. The Hamiltonian of the system is given by (we adopt a unit system with $\hbar=m=\omega=1$ where $m$ is the mass of the atom and $\omega$ the frequency of the harmonic trap):
\begin{equation}
        H= \sum_{i=1}^N\left(-\frac{1}{2}\frac{\partial^2}{\partial x_i^2}+\frac{1}{2} x_i^2\right) +\sum_{i>j}\sum_{J}g_{J}\hat{ P}^{ij}_{J} \delta(x_i-x_j)\,, \label{H}
\end{equation}
where $\hat{ P}^{ij}_{J}$ is the spin projection operator acting on particles $i,j$ into the coupled spin $J$ of the two particles, and $g_J$ is the interaction strength in this spin channel. The summation over $J$ is limited to symmetric (antisymmetric) channels for bosons (fermions). We will focus in this work on the hardcore limit with $g_J \rightarrow \infty$, but will comment on the effect of large but finite $g_J$ in the end.

The many-body wave function must satisfy the hardcore boundary condition: $\left. \Psi \right|_{x_i=x_j}=0$, i.e., the wave function vanishes when the coordinates of any two atoms, regardless of their spins, coincide with each other. This implies that, in any spatial sector (say, the sector denoted as $S_1$ with $x_1<x_2<...<x_N$), the wave function separates into a spatial and a spin part and the energy is independent of the spin configuration. The spatial part can be represented by a Slater determinant denoted as $\phi_F(x_1,x_2,...,x_N)$, which is an eigenstate of the ideal spinless Fermi gas, projected onto the sector $S_1$. The total many-body wave function then takes the following form \cite{Deuretzbacher2008, GuanL2009}:
\begin{equation}
    \Psi(x_1,\sigma_1;...;x_N,\sigma_N) = \sum_P (\pm 1)^P\,P \left[\phi_F \theta^1 \otimes \chi \right]\,,\label{Psi}
\end{equation}
where $P$ stands for permutation, $(\pm 1)$ refers bosons and fermions, respectively, $\theta^1=\Pi_{i=1}^{N-1} \theta(x_{i+1}-x_i)$, with $\theta(x)$ being the Heaviside step function, can be regarded as the projection operator onto the spatial sector $S_1$, and $\chi$ is an arbitrary wave function for a 1D spin chain. 

To find the real space density profile and momentum distribution associated with $\Psi$, we need to introduce the one-body density matrix (OBDM) which for a general many-body wave function is defined as:
\begin{widetext}
\begin{equation}
    \rho_{\sigma' \sigma} (x',x)=N\mathlarger{\sum}_{\sigma_2,...,\sigma_N} \int dx_2 ... dx_N\, \Psi^*(x',\sigma';x_2,\sigma_2;...;x_N,\sigma_N)\Psi(x,\sigma;x_2,\sigma_2;...;x_N,\sigma_N)\,.
\end{equation}
\end{widetext}
For the wave function given in (\ref{Psi}), it has been shown \cite{yang_strongly_2015} that the OBDM takes the following form:
\begin{equation}
    \rho_{\sigma'\sigma}(x',x)=\sum_{m,\,n=1}^N \,\rho_{mn}(x',x)S_{mn}(\sigma',\sigma)\,,
    \label{eq:OBDM}
\end{equation}
where 
\begin{align}
\rho_{mn}(x',x)=(-1)^{n-m}N!\int_{\Gamma_{mn}} \!\! dx_2...& dx_N \, \phi_F^*(x', x_2,...,x_N) \nonumber \\ & \;\;\;
\phi_F(x, x_2,...,x_N)\,, \label{rhox}
\end{align}
with $\Gamma_{mn}$ denoting an ordering such that $x_2<...<x_{m-1}<x'<x_m<...<x_{n-1}<x<x_n<..<x_N$, and the spin correlations are given by
\begin{equation}
    S_{mn}(\sigma',\sigma)=(\pm 1)^{n-m}\sum_{\sigma_2..\sigma_N}\braket{\chi|c^\dagger_m(\sigma') (m...n)c_n(\sigma)|\chi}\,,
\end{equation}
where $c_n(\sigma)$ annihilates a spin $\sigma$ at the $n^{\rm th}$ position, $(m...n)$ is a loop permutation operator that cyclically permutes particles and maps ~${\{m,..,n\}} \to {\{m+1,......,n-1,n\}}$. Given the OBDM, the real space density profile and the momentum distribution for spin component-$\sigma$ are respectively given by
\begin{align}
n_\sigma(x) &=  \rho_{\sigma \sigma}(x,x)= \sum_m \rho_{mm}(x,x)S_{mm}(\sigma,\sigma) \,, \label{nx}\\
\tilde{n}_\sigma(k) &= \frac{1}{2\pi} \int dx \int dx' \,e^{ik(x-x')}\,\rho_{\sigma \sigma}(x,x') \,. \label{np}
\end{align}
Both $n_\sigma(x)$ and $\tilde{n}_\sigma(k)$ depend on the spin configuration $\chi$. Using $\sum_\sigma S_{mm}(\sigma,\sigma)=1$, it is straightforward to show that the total density profile in real space 
\begin{equation}
    \sum_\sigma n_\sigma(x) =n_F(x)\label{eq:fermionization}\,,
\end{equation} 
is {\em independent} of the spin state $\chi$, where $n_F(x)$ is the density profile of the ideal Fermi gas whose wave function is given by $\phi_F$. This is the fermionization of the hardcore spinor quantum gas. 

We are now ready to prove the two properties mentioned earlier that are associated with DF. We assume that the spinor gas is initially prepared in its ground state in the presence of the harmonic trap, whose wave function takes the form of Eq.~(\ref{Psi}), where $\chi$ is arbitrary due to the spin degeneracy and $\phi_F$ is the Slater determinant constructed from the $N$ lowest-energy single-particle harmonic oscillator eigenstates $\phi_n=(2^n n!\sqrt{\pi})^{-1/2}H_n(x) e^{-x^2/2}$ ($n=0,1,...,N-1$), which we denote as
\begin{equation}
    \phi_F = {\rm Det}[\phi_0(x), \phi_1(x),...,\phi_{N-1}(x)  ]/\sqrt{N!}\,. \label{Det}
\end{equation}
At $t=0$ the trap is suddenly turned off. Crucially, due to the hardcore constraint, the spin configuration remains frozen. As a consequence, the spin correlation function $S_{mn}(\sigma',\sigma)$ in the OBDM (\ref{eq:OBDM}) does not evolve in time. The time dependence of the OBDM is carried by $\rho_{mn}(x',x)$, and hence by $\phi_F(t)$ according to Eq.~(\ref{rhox}). On the other hand, $\phi_F(t)$ is related to $\phi_F(0)$ as
\begin{align}
    \phi_F(x_1,x_2,...,x_N;t)=&b^{-N/2}\, \phi_F\left(\frac{x_1}{b},\frac{x_2}{b},..., \frac{x_N}{b};0\right) \nonumber \\
    &\exp{\left[ i\frac{\dot{b}}{b}\sum_{i=1}^N \left( \frac{x_i^2}{2}- E_{i-1}\tau(t) \right)\right]}\,,
    \label{eq:scale}
\end{align}
where $E_i$ is the energy of the $i^{\rm th}$ single-particle eigenstate of the initial harmonic trap, $b(t)=\sqrt{1+ t^2}$ is the spatial scaling parameter, and $\tau(t)=\int^t_0 dt' b^{-2}(t')$ the time scaling parameter. Equation (\ref{eq:scale}) follows from the scaling solution of the harmonic oscillator state under a parametric modulation of the trapping frequency \cite{RussianPaper1,RussianPaper2,kagan_evolution_1996}. From Eqs.~(\ref{eq:OBDM}), (\ref{rhox}) and (\ref{eq:scale}), it immediately follows that the OBDM at time $t$ is related to the initial OBDM as 
\begin{equation}
    \rho_{\sigma'\sigma}(x',x;t)=\frac{1}{b}\exp\left[\frac{i\dot{b}}{2b}(x^2-{x'}^2)\right]\rho_{\sigma'\sigma}(x'/b,x/b;0)\,. \label{obdmt}
\end{equation}
The real space density profile is immediately obtainable:
\begin{equation}
    n_{\sigma}(x;t)= \rho_{\sigma\sigma}(x,x;t)= \frac{1}{b}\,n_\sigma(x/b;0)\,,
\end{equation}
which describes a self-similar expansion for each spin component.

The evolution of the momentum distribution can be obtained by inserting Eq.~(\ref{obdmt}) into (\ref{np}). The integral in general does not yield closed form expression. However, exact results can be obtained in the following two limits. The first concerns the large momentum limit $|k| \rightarrow \infty$. Wave functions possessing cusp singularities would lead to a power law momentum tail $\tilde{n}_\sigma(k) \sim \kappa_\sigma/ k^{4}$, where the coefficient $\kappa_\sigma$ is the Tan contact for the $\sigma$ spin component. \cite{TAN20082952,BARTH20112544}. As has been argued by Minguzzi and Gangardt in their study of the DF for Tonks-Girardeau gas \cite{minguzzi_exact_2005}, the scaling behavior represented by Eqs.~(\ref{eq:scale}) and (\ref{obdmt}) means that the cusp structure of the wave function remains the same during the expansion. As a result, the momentum tail exhibits a similar scaling behavior:
\begin{equation}
    \tilde{n}_\sigma(k;t) \sim \frac{1}{b^3(t)}\, \frac{\kappa_\sigma}{k^4}\,,\;\;\;|k| \rightarrow \infty \label{tan}
\end{equation}
which indicates that the Tan contact decreases in time by a factor of $1/b^3$. The second limit where analytic results can be obtained is the asymptotic limit $t \rightarrow \infty$ (for which, $b \rightarrow t$ and $\dot{b} \rightarrow 1$), under which the integral can be greatly simplified by invoking the stationary phase approximation \cite{minguzzi_exact_2005} and we have
\begin{equation}
    \tilde{n}_\sigma(k;t \rightarrow \infty) = \rho_{\sigma \sigma}(k,k;0)=n_\sigma(k;0)\,,\label{nkt}
\end{equation}
which, according to Eq.~(\ref{nx}), has the same shape as the initial real space density profile inside the trap (\textbf{Property 2}). It is amusing to note that this is just the opposite situation of the ballistic expansion, under which the asymptotic real space density profile takes the shape of the initial momentum distribution in the trap. The asymptotic total momentum distribution follows from \textbf{Property 2} and the fermionization of the spinor gas (Eq.~\ref{eq:fermionization}):
\[\tilde{n}(k; t \rightarrow \infty) \equiv \sum_\sigma \tilde{n}_\sigma(k;t \rightarrow \infty)=n_F(k;0)\,,\] and therefore takes the shape of the initial total real space density profile, which is the same as the momentum distribution, $\tilde{n}_F(k)$, of the spinless Fermi gas in the trap (\textbf{Property 1}). 
We have thus succeeded in proving the two properties concerning DF of the hardcore spinor gas.

\begin{figure}
  \centering
    \includegraphics[width=0.5\textwidth]{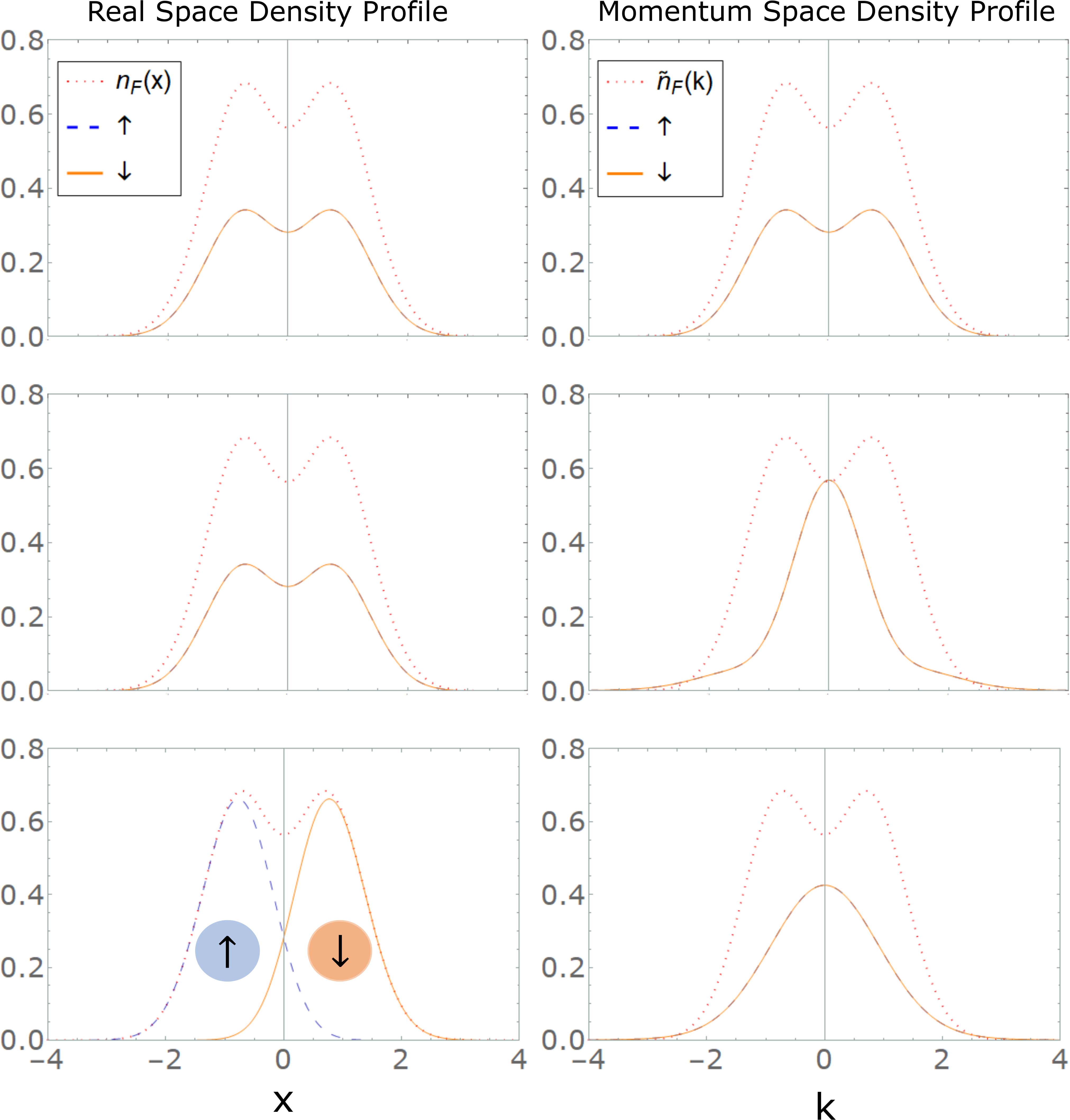}
      \caption{(color online) Real (left panel) and momentum (right panel) space density profiles of two spinor hardcore bosons. The top, middle, and bottom rows correspond to the states $\Psi^S$ [Eq.~(\ref{Psi1})], $\Psi^T$ [Eq.~(\ref{Psi2})], and $\Psi$ [Eq.~(\ref{psits})], respectively. The dashed and solid lines correspond to spin-$\uparrow$ and spin-$\downarrow$, respectively. The dotted lines correspond to two spinless fermions. 
  \label{fig:initial}}
\end{figure}

\emph{Two particle example --} We now consider a specific example of two hardcore bosons of different spins (denoted as $\uparrow$ and $\downarrow$). This simple case allows straightforward calculations, and in the same time yields all the essential features of DF for a general many-body spinor system. The trapped spinless fermion wave function of Eq.~(\ref{Det}) now takes the explicit form:
\begin{equation}
    \phi_F(x_1,x_2)=\pi^{-1/2}(x_2-x_1)\, e^{ -(x_1^2+x_2^2)/2 }\,.
\end{equation}
We can construct the following as a ground state wave function of the spin-1/2 hardcore bosons:
\begin{equation}
    \Psi^S =\phi_F(x_1,x_2) \otimes \chi^S\,, \label{Psi1}
\end{equation}
where $\chi^S = (|\uparrow \downarrow \rangle - |\downarrow \uparrow \rangle)/\sqrt{2} $ is the spin singlet state. $\Psi^S$ is a direct product of a spatial and a spin wave function. This separation of the two degrees of freedom makes the calculation of the OBDM rather simple. In fact, we can show that the OBDM for the two spin components are the same and given by:
\begin{equation}
    \rho^S_{\uparrow \uparrow}(x',x) =\rho^S_{\downarrow \downarrow}(x',x)= \rho_F(x',x)/2\,,
\end{equation}
where $\rho_F(x',x)$ is the OBDM for the spinless fermion. As a result, the real space density profile and the momentum distribution of the spin-1/2 system (shown in the top row of Fig.~\ref{fig:initial}) are simply determined by those of the spinless fermions. After the trap is turned off, the former will exhibit self-similar expansion, while the latter remains fixed in time.

We can construct another ground state wave function of the spin-1/2 system as:
\begin{equation}
    \Psi^T =\phi_B(x_1,x_2) \otimes \chi^T\,, \label{Psi2}
\end{equation}
where $\chi^S = (|\uparrow \downarrow \rangle + |\downarrow \uparrow \rangle)/\sqrt{2} $ is the triplet state, and $\phi_B$ is the wave function of the Tonks-Girardeau gas consisting of two hardcore spinless bosons. According to the Bose-Fermi mapping, we know that $\phi_B(x_1,x_2) = |\phi_F(x_1,x_2)|$. $\Psi^T$ is again a product state, and following a similar procedure as above, one can again show that the OBDM for the two spin components are the same and given by
\begin{equation}
    \rho^T_{\uparrow \uparrow}(x',x) =\rho^T_{\downarrow \downarrow}(x',x) =  \rho_B(x',x)/2\,,
\end{equation}
where $\rho_B(x',x)$ is the OBDM for the hardcore spinless bosons. Hence now the real space density profile and the momentum distribution (shown in the middle row of Fig.~\ref{fig:initial}) are identical to those of the Tonks-Girardeau gas.

Now let us consider a superposition of $\Psi^T$ and $\Psi^S$:
\begin{equation}
    \Psi = (\Psi^T + \Psi^S)/\sqrt{2} \,, \label{psits}
\end{equation}
which is no longer a product state of a spatial and a spin wave function but, due to the degeneracy of $\Psi^T$ and $\Psi^S$, is also a ground state of the spin-1/2 system. Writing in the form of Eq.~(\ref{Psi}), the corresponding spin-chain state is represented by $\chi =|\uparrow \downarrow \rangle$, with spin-up to the left of spin-down. The calculation of the OBDM is more involved than the previous case but also straightforward. One can show that
\begin{align}
\rho_{\uparrow \uparrow} (x',x)&= \frac{1}{2} \int dx_2 \,(\phi_B+\phi_F)(\phi'_B + \phi'_F) \,, \\
\rho_{\downarrow \downarrow} (x;,x) &= \frac{1}{2} \int dx_2 \,(\phi_B-\phi_F)(\phi'_B - \phi'_F)\,,
\end{align}
where we have introduced the short-hand notation: $\phi_B \equiv \phi_B(x,x_2)$, $\phi'_B \equiv \phi_B(x',x_2)$, and similarly for $\phi_F$ and $\phi'_F$. The real space density profile inside the trap can be found easily as
\begin{equation}
    n_{\uparrow, \downarrow}(x) = [n_F(x) \pm n_C(x)]/2\,,
\end{equation}
where $n_F(x) = e^{-x^2}(1+2x^2)/\sqrt{\pi}$ is the density profile of the spinless fermion, and 
\begin{align*}
n_C(x) = 2\!\int \! dx_2 \phi_B \phi_F = -\frac{1}{\pi} \!\left[ 2xe^{-2x^2} \!\!\!+\!e^{-x^2}(1+2x^2) {\rm erf}(x) \right].
\end{align*}
The density profiles are plotted in the left panel of the bottom row in Fig.~\ref{fig:initial}, which shows that spin-up (spin-down) atom occupies the left (right) side of the trap. 

The momentum distribution can also be readily calculated, and one can show that the two spin components possess identical initial momentum distribution given by
\begin{equation}
    \tilde{n}_{\uparrow , \downarrow} (k)= \left[\tilde{n}_F(k) +\tilde{n}_B(k) \right]/4\,,
\end{equation}
and are shown in the right panel of the bottom row in Fig.~\ref{fig:initial}.

\begin{figure}
  \centering
    \includegraphics[width=0.5\textwidth]{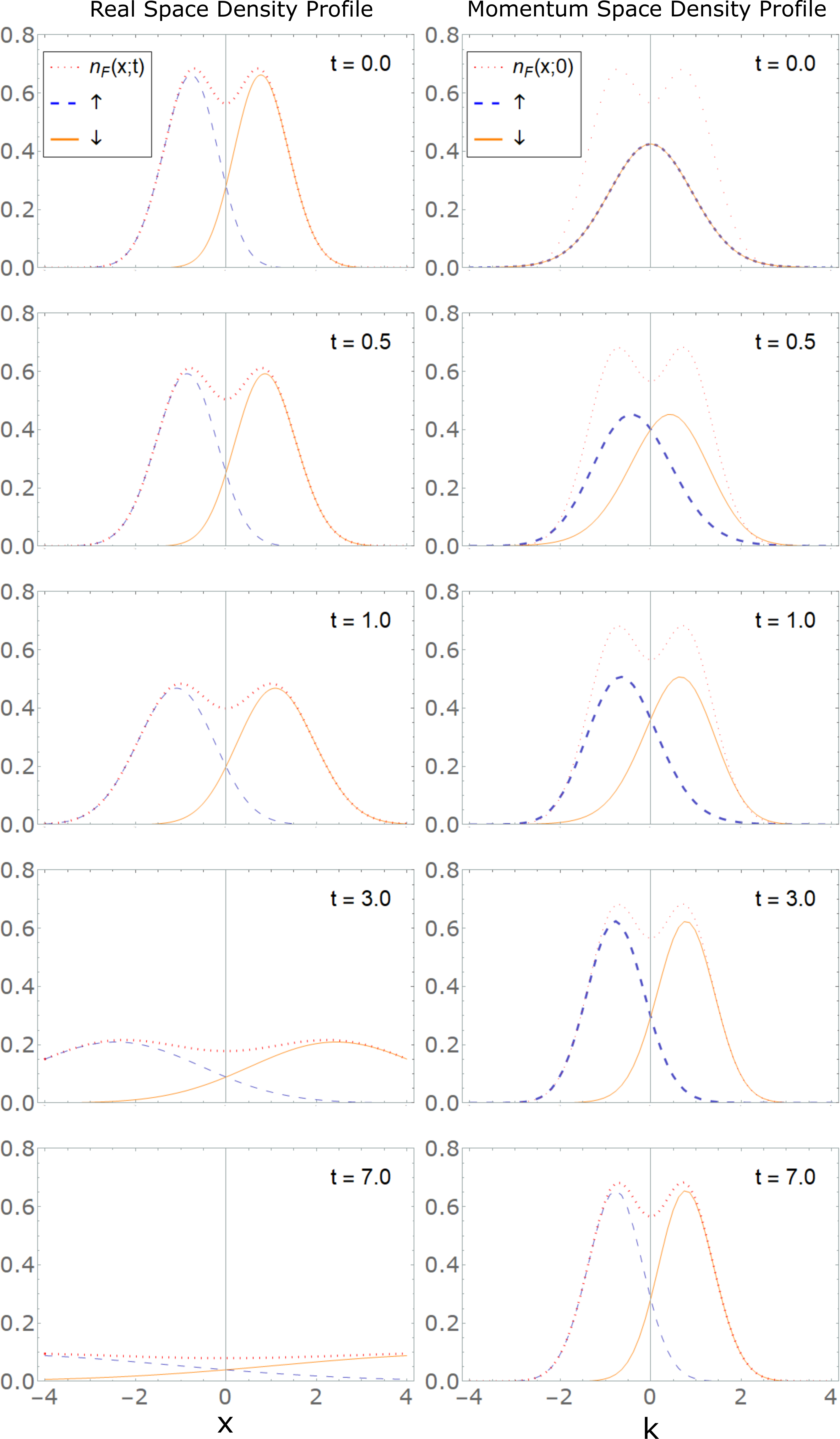}
      \caption{(color online) Real space and momentum space distributions of two spinor hardcore bosons, initially prepared in state $\Psi$, as in Eq.~(\ref{psits}), after quench.
  \label{fig:finalplot}}
\end{figure}

To investigate the time evolution after the quench of the trapping potential, particularly that of the momentum distribution, we need to resort to numerics to work out the integrals. We present the results in Fig.~\ref{fig:finalplot}. The left column of Fig.~\ref{fig:finalplot} displays the evolution of the real space density profile, which simply goes through a self-similar expansion for both spins. The right column of Fig.~\ref{fig:finalplot} shows the evolution of the momentum distribution, whose shape asymptotically approaches the initial real space density profile.  

Using the method outlined in Ref.~\cite{Olshanii2003}, the momentum tail at time $t$ can be explicitly found as
\begin{equation}
    \lim_{|k| \rightarrow \infty} \tilde{n}_{\uparrow}(k;t) = \tilde{n}_{\downarrow}(k;t) =\frac{1}{b^3(t)} \left( \frac{2}{\pi} \right)^{3/2} \frac{1}{4k^4}\,,\label{tail}
\end{equation}
consistent with the qualitative argument given in Eq.~(\ref{tan}). Note that, even though $\tilde{n}_{\uparrow, \downarrow}(k;t>0)$ are not symmetric about $k=0$, they do have symmetric momentum tails.

{\em Away from the hardcore limit for $N$ particles ---} So far we have focused solely on the hardcore limit. Let us now turn to discuss the effect of large but finite interaction strengths. For finite $g_J$, the spin degeneracy is lifted. To the leading order, the charge degrees of freedom remain frozen and the system is still fermionized i.e., the total density profile coincides with that of an ideal spinless Fermi gas. However, the spin degrees of freedom are governed by an effective spin-chain Hamiltonian 
\begin{equation}
    H_{\mathrm{sc}}= -\sum_{i=1}^{N-1} C_i \sum_J \frac{\hat{P}^{J}_{i,i+1}}{g_J}\,, \label{sc}
\end{equation}
where $\hat{P}^{J}_{i,i+1}$ is the exchange operator for the two neighboring spins at the $i^{\rm th}$ and the $(i+1)^{\rm th}$ positions when their total spin is $J$, and the exchange coefficients $C_i$ are given by:
\begin{equation}
    C_i=2N! \int dx_1...dx_N\, \left|\frac{\partial \phi_F}{\partial x_i}\right|^2 \delta (x_i - x_{i+1})\theta^1_{[i+1,i]}\,,\label{ci}
\end{equation}
where $\theta^1_{[i+1,i]}=\theta^1/\theta(x_{i+1}-x_i)$ is a reduced Heaviside function.
One remarkable aid in exploring strongly interacting spinor gases is that the wave function is still in the form of Eq.~(\ref{Psi}), but here the spin wave function $\chi$ is no longer arbitrary and is instead determined by the effective spin-chain Hamiltonian in Eq. (\ref{sc})~\cite{yang_strongly_2015}. 

In studying the DF for hardcore spinor gases, one crucial point is that the spin configuration characterized by $\chi$ remains fixed after the trap is turned off. One might expect that this would no longer be the case for finite $g_J$, as now $\chi$ should evolve under Hamiltonian (\ref{sc}) which itself is time-dependent due to the time-dependence of $\phi_F$ which will determine the coefficients $C_i$, see Eq.~(\ref{ci}). Remarkably, for the initial harmonic trap, the scaling behavior of $\phi_F$ in Eq.~(\ref{eq:scale}) leads to a scaling behavior of $C_i$:
$ C_i(t) = C_i(0)b^{-3}(t)$.
As a result, the effective spin-chain Hamiltonian simply undergoes an overall scaling \cite{volosniev_simulation_2016}: 
\[ H_{\rm sc}(t) = \frac{1}{b^3(t)} H_{\rm sc}(0) \,.\]
Hence an eigenstate of the initial Hamiltonian $H_{\rm sc}(0)$ remains as an eigenstate of $H_{\rm sc}(t)$ for later time. In other words, we still have the important observation that the spin configuration does not change during the cloud expansion. Therefore, the DF properties for hardcore systems remain valid in the strongly interacting limit. We emphasize that this freezing of the spin configuration during expansion for finite $g_J$ is a special property for harmonic traps, and is not expected to be valid for other trapping potentials. The expansion dynamics for finite $g_J$ and for initial non-harmonic trapping potential is interesting problem and will be studied in near future.

{\em Conclusion ---} We have provided an analytical proof of the DF phenomenon for a general hardcore spinor gas --- either fermions or bosons with arbitrary spin --- initially confined in a harmonic trap. After the trap is turned off, the total momentum distribution approaches that of a spinless ideal Fermi gas, while the asymptotic momentum distribution of each spin component takes the same shape of the initial real space density profile of that spin component. Furthermore, we have argued that these properties remain true when the system is slightly away from the hardcore limit. Our work helps us to better understand the interplay between the charge and the spin degrees of freedom of a spinor gas. In a bigger context,  coherent quantum dynamics is a very active frontier of quantum many-body research. The phenomena we studied here represents a precious example where such quantum dynamics can be investigated and understood analytically, as such it can serve as a benchmark for more general studies.

We would like to thank Drs. Xiwen Guan and David Weiss for many useful discussions. This work is supported by the NSF and the Welch Foundation (Grant No. C-1669). This work was done by L. Yang and T. Skaras at Rice University; they are now at Google and Cornell University respectively.

\bibliography{mybib}

\end{document}